\documentclass[10pt]{article}
\usepackage{latexsym,graphicx}
\newcommand{\be}{\begin{equation}}
\newcommand{\ee}{\end{equation}}
\def\n{\noindent}
\catcode `\@=11
\catcode `\@=12
\begin{document}
\begin{center}
\large{\bf {Some Bianchi Type III String Cosmological Models with Bulk Viscosity}} \\
\vspace{10mm}
\normalsize{Mahesh Kumar Yadav $^1$, Anju Rai $^2$ and Anirudh Pradhan$^3$
 \\
\vspace{5mm}
\normalsize{$^{1}$Department of Mathematics, Sobhasaria Engineering College,
NH - 11, Sikar-332 001, India \\
E-mail : yadavmk@iucaa.ernet.in}\\
\vspace{5mm}
\normalsize{$^{2,3}$Department of Mathematics, Hindu Post-graduate College, 
Zamania-232 331, Ghazipur, India \\
$^{2}$E-mail : anjuraianju@yahoo.co.in}} \\
$^{3}$E-mail : pradhan@iucaa.ernet.in, acpradhan@yahoo.com\\
\vspace{5mm}
\vspace{5mm}
\end{center}
\vspace{10mm}
\begin{abstract} 
We investigate the integrability of cosmic strings in Bianchi III space-time in 
presence of a bulk viscous fluid by applying a new technique. The behaviour of the 
model is reduced to the solution of a single second order nonlinear differential 
equation. We show that this equation admits an infinite family of solutions. Some
physical consequences from these results are also discussed. 
\end{abstract}
 \smallskip
\n Keywords : Cosmic string, viscous models, integrable \\
\n PACS number: 98.80.Cq, 04.20.-q 
\section{Introduction}
In recent years, there has been considerable interest in string cosmology. Cosmic 
strings are topologically stable objects which might be found during a phase 
transition in the early universe (Kibble, 1976). Cosmic strings play an important role in 
the study of the early universe. These arise during the phase transition after the 
big bang explosion as the temperature goes down below some critical temperature as 
predicted by grand unified theories (Zel'dovich et al., 1975; Kibble, 1976, 1980; 
Everett, 1981; Vilenkin, 1981). It is believed that cosmic strings give rise to density 
perturbations which lead to the formation of galaxies (Zel'dovich, 1980). These cosmic 
strings have stress-energy and couple to the gravitational field, it may be interesting to 
study the gravitational effects that arise from strings. 

The general relativistic treatment of strings was initiated by Letelier (1979, 1983) and 
Stachel (1980). Letelier (1979) has obtained the solution to Einstein's field equations 
for a cloud of strings with spherical, plane and cylindrical symmetry. Then, in 1983, he 
solved Einstein's field equations for a cloud of massive strings and obtained cosmological 
models in Bianchi I and Kantowski-Sachs space-times. Benerjee et al. (1990) have investigated 
an axially symmetric Bianchi type I string dust cosmological model in presence and absence 
of magnetic field. The string cosmological models with a magnetic field are also discussed 
by Chakraborty (1991), Tikekar and Patel (1992, 1994). Patel and Maharaj (1996) investigated 
stationary rotating world model with magnetic field. Ram and Singh (1995) obtained some new exact 
solutions of string cosmology with and without a source free magnetic field for Bianchi type I 
space-time in the different basic form considered by Carminati and McIntosh (1980). Exact 
solutions of string cosmology for Bianchi type II, $VI_0$, VIII and IX space-times have been 
studied by Krori et al. (1990) and Wang (2003). Singh and Singh (1999) investigated string 
cosmological models with magnetic field in the context of space-time with $G_{3}$ 
symmetry. Singh (1995) has also studied string cosmological models with electromagnetic field 
in Bianchi type II, VIII and IX space-times. Lidsey, Wands and Copeland (2000) have reviewed 
aspects of super string cosmology with the emphasis on the cosmological implications of duality 
symmetries in the theory. Baysal et al. (2001) have investigated the behaviour of a string in the 
cylindrically symmetric inhomogeneous universe. Bali et al. (2001, 2003, 2006) have 
obtained Bianchi types I and  IX string cosmological models in general relativity. Yavuz 
(2005) have examined charged strange quark matter attached to the string cloud in the 
spherical symmetric space-time admitting one-parameter group of conformal motion. Recently 
Kaluza-Klein cosmological solutions are obtained by Yilmaz (2006) for quark matter coupled 
to the string cloud in the context of general relativity.

On the other hand, the matter distribution is satisfactorily described by perfect fluids due 
to the large scale distribution of galaxies in our universe. However, a realistic treatment 
of the problem requires the consideration of material distribution other than the perfect 
fluid. It is well known that when neutrino decoupling occurs, the matter behaves as a viscous 
fluid in an early stage of the universe. Viscous fluid cosmological models of early universe 
have been widely discussed in the literature.   

Tikekar and Patel (1992), following the techniques used by Letelier and Stachel, obtained 
some exact Bianchi III cosmological solutions of massive strings in presence of magnetic 
field. Maharaj et al. (1995) have generalized the previous solutions obtained by Tikekar 
and Patel (1992) by considering Lie point symmetries. Recently Yadav et al. (2006) have 
studied some Bianchi type I viscous fluid string cosmological models with magnetic field. 
Recently Wang (2003, 2004, 2005, 2006) has discussed LRS Bianchi type I and Bianchi type III 
cosmological models for a cloud string with bulk viscosity. Motivated the situations 
discussed above, in this paper, we shall focus upon the problem of establishing a formalism 
for studying the new integrability of cosmic strings in Bianchi III space-time in presence of 
a bulk viscous fluid by applying a new technique.

\section{The Metric and Field  Equations}
We consider the space-time of general Bianchi III type with the metric 
\begin{equation}
\label{eq1}
ds^{2} =  - dt^{2} + A^{2}(t) dx^{2} + B^{2}(t) e^{-2ax} dy^{2} + C^{2}(t) dz^{2},
\end{equation}
where $a$ is constant. The energy momentum tensor for a cloud of string dust 
with a bulk viscous fluid of string is given by Letelier (1979) and Landau \& 
Lifshitz (1963) 
\begin{equation}
\label{eq2}
T^{j}_{i} = \rho v_{i}v^{j} - \lambda x_{i}x^{j} - \xi v^{l}_{;l} \left(g^{j}_{i} 
+ v_{i} v^{j}\right),
\end{equation}
where $v_{i}$ and $x_{i}$ satisfy condition
\begin{equation}
\label{eq3}
v^{i} v_{i} = - x^{i} x_{i} = -1, \, \, \, v^{i} x_{i} = 0.
\end{equation}
In equations (\ref{eq2}) $\rho$ is the proper energy density for a cloud string with 
particles attached to them, $\lambda$ is the string tension density, $v^{i}$ is the 
four-velocity of the particles and $x^{i}$ is a unit space-like vector representing 
the direction of string. If the particle density of the configuration is denoted by 
$\rho_{p}$, then we have
\begin{equation}
\label{eq4}
\rho = \rho_{p} + \lambda.
\end{equation}
The Einstein's field equations (in gravitational units $c = 1$, $G = 1$) read as
\begin{equation}
\label{eq5}
R^{j}_{i} - \frac{1}{2} R g^{j}_{i} = - 8\pi T^{j}_{i},
\end{equation}
where $R^{j}_{i}$ is the Ricci tensor; $R$ = $g^{ij} R_{ij}$ is the
Ricci scalar. In a co-moving co-ordinate system, we have
\begin{equation}
\label{eq6}
v^{i} = (0, 0, 0, 1), \, \, \, x^{i} = (0, 0, 1/C, 0).
\end{equation}
The field equations (\ref{eq5}) with (\ref{eq2}) subsequently lead to the following 
system of equations:
\begin{equation}
\label{eq7}
\frac{B_{44}}{B} + \frac{C_{44}}{C} + \frac{B_{4} C_{4}}{B C} = 8\pi \xi \theta,
\end{equation}
\begin{equation}
\label{eq8}
\frac{A_{44}}{A} + \frac{C_{44}}{C} + \frac{A_{4} C_{4}}{A C} = 8\pi \xi \theta,
\end{equation}
\begin{equation}
\label{eq9}
\frac{A_{44}}{A} + \frac{B_{44}}{B} + \frac{A_{4} B_{4}}{A B} - \frac{a^{2}}{A^{2}} = 
8\pi (\lambda + \xi \theta),
\end{equation}
\begin{equation}
\label{eq10}
\frac{A_{4}B_{4}}{A B} + \frac{B_{4} C_{4}}{B C} + \frac{C_{4} A_{4}}{C A} - \frac{a^{2}}
{A^{2}}  = 8\pi \rho,
\end{equation}
\begin{equation}
\label{eq11}
\frac{A_{4}}{A} - \frac{B_{4}}{B} = 0,
\end{equation}
where the suffix $4$ at the symbols $A$, $B$ and $C$ denotes ordinary differentiation 
with respect to $t$.
The particle density $\rho_{p}$ is given by 
\begin{equation}
\label{eq12}
8\pi \rho_{p} = \frac{B_{4} C_{4}}{B C} + \frac{C_{4} A_{4}}{C A} - \frac{B_{44}}{B} 
+ \frac{C_{44}}{C} + \frac{A_{4} C_{4}}{A C} 
\end{equation}
in accordance with (\ref{eq4}). \\

The velocity field $v^{i}$ as specified by (\ref{eq6}) is irrotational. The scalar 
expansion $\theta$ and components of shear $\sigma_{ij}$ are given by
\begin{equation}
\label{eq13}
\theta = \frac{A_{4}}{A} + \frac{B_{4}}{B} + \frac{C_{4}}{C},
\end{equation}
\begin{equation}
\label{eq14}
\sigma_{11} = \frac{A^{2}}{3}\left[\frac{2A_{4}}{A} - \frac{B_{4}}{B} - \frac{C_{4}}{C}\right],
\end{equation}
\begin{equation}
\label{eq15}
\sigma_{22} = \frac{B^{2} e^{-2ax}}{3}\left[\frac{2B_{4}}{B} - \frac{A_{4}}{A} - \frac{C_{4}}{C}
\right],
\end{equation}
\begin{equation}
\label{eq16}
\sigma_{33} = \frac{C^{2}}{3}\left[\frac{2C_{4}}{C} - \frac{A_{4}}{A} - \frac{B_{4}}{B}\right],
\end{equation}
\begin{equation}
\label{eq17}
\sigma_{44} = 0.
\end{equation}
Therefore
\begin{equation}
\label{eq18}
\sigma^{2} = \frac{1}{3}\left[\frac{A_{4}^{2}}{A^{2}} + \frac{B_{4}^{2}}{B^{2}} + 
\frac{C_{4}^{2}}{C^{2}} - \frac{A_{4}B_{4}}{AB} - \frac{B_{4}C_{4}}{BC} - 
\frac{C_{4}A_{4}}{CA}\right]. 
\end{equation}
\section{Solutions of the Field  Equations}
The field equations (\ref{eq7})-(\ref{eq11}) are a system of five equations with six 
unknown parameters $A$, $B$, $C$, $\rho$, $\lambda$ and $\xi$. One additional constraint 
relating these parameters is required to obtain explicit solutions of the system. We 
assume that the expansion ($\theta$) in the model is proportional to the eigen value 
$\sigma^{2}_{2}$ of the shear tensor $\sigma^{j}_{i}$. This condition leads to 
\begin{equation}
\label{eq19}
B = \alpha (A C)^{\beta},
\end{equation}
where $\alpha$ and $\beta$ are arbitrary constants. Equations (\ref{eq11}) leads to
\begin{equation}
\label{eq20}
A = m B,
\end{equation}
where $m$ is an integrating constant.
From Eqs. (\ref{eq19}) and (\ref{eq20}), we obtain
\begin{equation}
\label{eq21}
B = M C^{N},
\end{equation}
where
\begin{equation}
\label{eq22}
M = \alpha^{\frac{1}{1 - \beta}} m^{\frac{\beta}{1 - \beta}}, \, \, \, N = \frac{\beta}
{1 - \beta}. 
\end{equation}
By the use of Eq. (\ref{eq20}) in field equations (\ref{eq8})-(\ref{eq10}) reduce to 
\begin{equation}
\label{eq23}
\frac{2B_{44}}{B} + \frac{B_{4}^{2}}{B^{2}} - \frac{a^{2}}{m^{2} B^{2}} = 8\pi  
(\lambda + \xi \theta),
\end{equation}
\begin{equation}
\label{eq24}
\frac{B_{4}^{2}}{B^{2}} + \frac{B_{4} C_{4}}{B C} - \frac{a^{2}}{m^{2} B^{2}} = 8\pi \rho.
\end{equation}
Using Eqs. (\ref{eq13}) and (\ref{eq21}) in (\ref{eq7}), we have
\begin{equation}
\label{eq25}
(N + 1)\frac{C_{44}}{C} + N^{2}\frac{C_{4}^{2}}{C^{2}} = 8\pi \xi (2N + 1)\frac{C_{4}}{C}.
\end{equation}
Let us consider 
\begin{equation}
\label{eq26}
C_{4} = f(C).
\end{equation}
Using Eq. (\ref{eq26}) in (\ref{eq25}), we get
\begin{equation}
\label{eq27}
\frac{df}{dC} + \left \{\left(\frac{N^{2}}{N + 1}\right)\frac{1}{C}\right\}f = 8\pi \xi
\left(\frac{2N + 1}{N + 1}\right).
\end{equation}
After integration, Eq. (\ref{eq27}), reduces to
\begin{equation}
\label{eq28}
f = 8\pi \xi \frac{(2N + 1)}{(N^{2} + N + 1)}C + \frac {P}{C\left(\frac{N^{2}}{N + 1}\right)},
\end{equation}
where $P$ is an integrating constant. Integrating (\ref{eq28}), we obtain
\begin{equation}
\label{eq29}
 C = \frac{1}{\xi^{k_{4}}}\left[k_{1} + k_{2} e^{k_{3} \xi t}\right]^{k_{4}},
\end{equation}
where $S$ is an integrating constant. Therefore
\begin{equation}
\label{eq30}
B = \frac{M}{\xi^{k_{5}}}\left [k_{1} + k_{2} e^{k_{3} \xi t}\right]^{k_{5}},
\end{equation}
\begin{equation}
\label{eq31}
A = \frac{m M}{\xi^{k_{5}}} \left[k_{1} + k_{2} e^{k_{3} \xi t}\right]^{k_{5}},
\end{equation}
where
$$k_{1} = -\frac{P(N^{2} + N + 1)}{8\pi (2N + 1)},$$
$$k_{2} =  \frac{S}{8\pi (2N + 1)},$$
$$k_{3} =  \frac{8\pi (2N + 1)}{(N + 1)},$$
$$k_{4} =  \frac{(N + 1)}{N^{2} + N + 1},$$
\begin{equation}
\label{eq32}
k_{5} =  \frac{N(N + 1)}{N^{2} + N + 1}.
\end{equation}
Hence the metric (\ref{eq1}) reduces to the form
\[
ds^{2} = - dt^{2} +  m^{2} M^{2}\left[\frac{k_{1} + k_{2} e^{k_{3} \xi t}}
{\xi}\right]^{2k_{5}} dx^{2} + M^{2} e^{-2a x}\left[\frac{k_{1} + k_{2} e^{k_{3} \xi t}}
{\xi}\right]^{2k_{5}}dy^{2}
\]
\begin{equation}
\label{eq33}
+ \left[\frac{k_{1} + k_{2} e^{k_{3} \xi t}}{\xi}\right]^{2k_{4}} dz^{2}.
\end{equation}
Using the suitable transformation
$$\frac{k_{1} + k_{2} e^{k_{3} \xi t}}{\xi} = L\frac{\sin{(\xi \tau)}}{\xi},$$
$$m M L^{k_{5}} x = X,$$
$$M L^{k_{5}} y = Y,$$
\begin{equation}
\label{eq34}
L^{k_{4}} z = Z, 
\end{equation}
the metric (\ref{eq33}) reduces to 
\[
ds^{2} = -\left(\frac{L\cos{(\xi \tau)}}{k_{3}\{k_{1} - L\sin{(\xi \tau)}\}}\right)^{2}d\tau^{2} + 
\left(\frac{\sin{(\xi \tau)}}{\xi}\right)^{2k_{5}} dX^{2} +
\] 
\begin{equation}
\label{eq35}
e^{-\frac{2 a X}{m M L^{k_{5}}}}\left(\frac{\sin{(\xi \tau)}}{\xi}\right)^{2k_{5}} dY^{2} +
\left(\frac{\sin{(\xi \tau)}}{\xi}\right)^{2k_{4}} dZ^{2}
\end{equation}
The rest energy ($\rho$), the string tension density ($\lambda$), the particle density 
($\rho_{p}$), expansion ($\theta$) and shear ($\sigma$) for the model (\ref{eq35}) are given by
\begin{equation}
\label{eq36}
8\pi \rho = k_{3}^{2}k_{5}(2k_{4} + k_{5})\left[\xi - \frac{k_{1}\xi}{L\sin{(\xi \tau)}}
\right]^{2} - \left(\frac{a}{mM}\right)^{2}\left[\frac{\xi}{L\sin{(\xi \tau)}}\right]^{2k_{5}},
\end{equation}
\[
8\pi \lambda = - 8\pi k_{3}\xi (k_{4} + 2k_{5})\left[\xi - \frac{k_{1}\xi}{L\sin{(\xi \tau)}}
\right] + 3k_{3}^{2} k_{5}^{2}\left[\xi - \frac{k_{1}\xi}{L\sin{(\xi \tau)}}
\right]^{2}
\]
\begin{equation}
\label{eq37}
+ 2 k_{1} k_{3}^{2} k_{5}\left[\frac{\xi^{2}}{{L\sin{(\xi \tau)}}} - 
\frac{k_{1}\xi^{2}}{L^{2}\sin^{2}{(\xi \tau)}}\right] - \left(\frac{a}{mM}\right)^{2}\left[\frac{\xi}
{L\sin{(\xi \tau)}}\right]^{2k_{5}},
\end{equation}
\[
8\pi \rho_{p} = 2k_{3}^{2} k_{5}(k_{4} - k_{5})\left[\xi - \frac{k_{1}\xi}
{L\sin{(\xi \tau)}}\right]^{2}
\]
\begin{equation}
\label{eq38}
+ \left[8\pi \xi k_{3}(k_{4} + 2k_{5}) - 2k_{1} k_{3}^{2} k_{5}\frac{\xi}
{L \sin{(\xi \tau)}}\right]\left[\xi - \frac{k_{1}\xi}{L\sin{(\xi \tau)}}\right],
\end{equation}
\begin{equation}
\label{eq39}
\sigma^{1}~ ~_{1} = \sigma^{2}~ ~_{2} = \frac{1}{3}k_{3}(k_{5} - k_{4})\left[\xi - 
\frac{k_{1}\xi}{L\sin{(\xi \tau)}}\right],
\end{equation}
\begin{equation}
\label{eq40}
\sigma^{3}~ ~_{3} =  \frac{2}{3}k_{3}(k_{4} - k_{5})\left[\xi - \frac{k_{1}\xi}
{L\sin{(\xi \tau)}}\right],
\end{equation}
\begin{equation}
\label{eq41}
\sigma^{4}~ ~_{4} =  0,
\end{equation}
\begin{equation}
\label{eq42}
\sigma^{2} =  \frac{1}{3}\Big[k_{3}(k_{5} - k_{4})\left(\xi - \frac{k_{1}\xi}
{L\sin{(\xi \tau)}}\right)\Big]^{2},
\end{equation}
\begin{equation}
\label{eq43}
\theta = k_{3}(2k_{5} + k_{4})\left[\xi - \frac{k_{1}\xi}{L\sin{(\xi \tau)}}\right].
\end{equation}

From Eqs. (\ref{eq36}) and (\ref{eq38}), we observe that the energy conditions 
$\rho \geq 0$ and $\rho_{p} \geq 0$ are fulfilled, provided
$$k_{5}(2k_{4} + k_{5})\left[\xi - \frac{k_{1}\xi}{L\sin{(\xi \tau)}}
\right]^{2} \geq \left(\frac{a}{k_{3}m M}\right)^{2}\left[\frac{\xi}
{L\sin{(\xi \tau)}}\right]^{2k_{5}},$$
and
$$ k_{3}^{2} k_{5}(k_{4} - k_{5})\left[\xi - \frac{k_{1}\xi}{L\sin{(\xi \tau)}}\right]^{2}
\geq $$
$$\left[k_{1} k_{3}^{2} k_{5}\frac{\xi}{L \sin{(\xi \tau)}} - 4\pi \xi k_{3}(k_{4} 
+ 2k_{5})\right] \left[\xi - \frac{k_{1}\xi}
{L\sin{(\xi \tau)}}\right],$$
respectively. 
From Eq. (\ref{eq37}), we observe that the string tension density $\lambda > 0$ provided
$$3k_{3}^{2} k_{5}^{2}\left[\xi - \frac{k_{1}\xi}{L\sin{(\xi \tau)}}\right]^{2} + 2 k_{1} k_{3}^{2} 
k_{5}\left[\frac{\xi^{2}}{{L\sin{(\xi \tau)}}} - \frac{k_{1}\xi^{2}}{L^{2}\sin^{2}
{(\xi \tau)}}\right] > $$
$$ \left(\frac{a}{mM}\right)^{2}\left[\frac{\xi}{L\sin{(\xi \tau)}}\right]^{2k_{5}} + 
8\pi k_{3}\xi (k_{4} + 2k_{5})\left[\xi - \frac{k_{1}\xi}{L\sin{(\xi \tau)}}\right]. $$

The model (\ref{eq35}) represents an expanding universe when $\sin {(\xi \tau)} > \frac{k_{1}}{L}$. 
When $\sin {(\xi \tau)} < \frac{k_{1}}{L}$, then $\theta$ decreases with time. Therefore the model 
describes a shearing non-rotating expanding universe without the big-bang start. We can see from the 
above discussion that the bulk viscosity plays a significant role in the evolution of the universe. 
Furthermore, since $\lim_{\tau \to \infty} \frac{\sigma}{\theta} \ne 0$, the model does not approach 
isotropy for large value of $\tau$. However, if $\sin {(\xi \tau)} = \frac{k_{1}}{L}$, the model 
(\ref{eq35}) represents an isotropic model in presence of bulk viscosity. 

In absence of bulk viscosity, i.e., when $\xi \to 0$, the metric (\ref{eq35}) 
reduces to
\begin{equation}
\label{eq44}
ds^{2} = -\left(\frac{L}{k_{1} k_{3}}\right)^{2}d\tau^{2} + (\tau)^{2k_{5}}dX^{2} + 
e^{-\frac{2aX}{m M L^{k_{5}}}}(\tau)^{2k_{5}} dY^{2} + (\tau)^{2k_{4}} dZ^{2}.
\end{equation}
The physical parameters $\rho$, $\lambda$, $\rho_{p}$ and the kinematical parameters $\theta$, 
$\sigma^{2}$ for this model are respectively given by 
\begin{equation}
\label{eq45}
8\pi \rho = \left(\frac{k_{1} k_{3}}{L \tau}\right)^{2}k_{5}(2 k_{4} + k_{5})  - 
\left(\frac{a}{mM}\right)^{2} \frac{1}{(L\tau)^{2k_{5}}},
\end{equation}
\begin{equation}
\label{eq46}
8\pi \lambda = \left(\frac{k_{1} k_{3}}{L\tau}\right)^{2}k_{5}(3k_{5} - 2) - 
\left(\frac{a}{mM}\right)^{2} \frac{1}{(L\tau)^{2k_{5}}},  
\end{equation}
\begin{equation}
\label{eq47}
8\pi \rho_{p} = \left[\frac{2k_{1}^{2} k_{3}^{2} k_{5}(1 + k_{4} - k_{5})}{L^{2}\tau^{2}}\right], 
\end{equation}
\begin{equation}
\label{eq48}
\sigma^{1}~ ~_{1} = \sigma^{2}~ ~_{2} = \frac{k_{1}k_{3}(k_{4} - k_{5})}{3L\tau},
\end{equation}
\begin{equation}
\label{eq49}
\sigma^{3}~ ~_{3} = \frac{2 k_{1}k_{3}(k_{5} - k_{4})}{3L\tau},
\end{equation}
\begin{equation}
\label{eq50}
\sigma^{4}~ ~_{4} = 0, 
\end{equation}
\begin{equation}
\label{eq51}
\sigma^{2} = \frac{1}{3}\left[\frac{k_{1}k_{3}(k_{4} - k_{5})}{L\tau}\right]^{2}, 
\end{equation}
\begin{equation}
\label{eq52}
\theta = - \frac{ k_{1}k_{3}(k_{4} + 2 k_{5})}{L\tau}. 
\end{equation}

From Eqs. (\ref{eq45}) and (\ref{eq47}), we observe that the energy conditions 
$\rho \geq 0$ and $\rho_{p} \geq 0$ are fulfilled, provided
$$k_{5}(2k_{4} + k_{5})(L \tau)^{2(k_{5} - 1)} \geq \left(\frac{a}{m M k_{1}k_{3}}\right)$$
and
$$ k_{4} \geq k_{5} - 1 $$
respectively. From Eq. (\ref{eq46}), we observe that the string tension density $\lambda \geq 0$ provided 
$$k_{5}(3k_{5} - 2)(L \tau)^{2(k_{5} - 1)} \geq \left(\frac{a}{m M k_{1}k_{3}}\right)^{2}.$$

In absence of bulk viscosity, the model (\ref{eq44}) starts expanding with a big bang at $\tau = 0$ 
and the expansion in the model decreases as time increases when $L , 0$. Also when $\tau \to \infty$ 
then shear is zero. Near the singularity $\tau = 0$, the physical parameters 
$\rho$, $\lambda$, $\rho_{p}$ are infinite if $k_{5} < 0$. Also since 
$\lim_{\tau \to \infty} \frac{\sigma}{\theta} \ne 0$, the model does not approach isotropy 
for large value of $\tau$. 

\section{Other Model}

In general, $\xi$ is not constant through out the fluid, so that $\xi$ can not be taken 
always constant, specially when the universe is expanding. Since, in general, $\xi$ depends 
on temperature $(T)$ and pressure$(p)$, it is reasonable to consider $\xi$ as function of 
time $t$.

In this case Eq. (\ref{eq25}), after integration, leads to
\begin{equation}
\label{eq53}
C = \left[b_{0} + k_{4}^{-1}\int{h(t)}dt\right]^{k_{4}},
\end{equation} 
where 
\begin{equation}
\label{eq54}
h(t) = c_{0} e^{k_{3}\int{\xi(t)}dt}.
\end{equation}
and $b_{0}$, $c_{0}$ are constants of integration.  
Therefore, we also obtain
\begin{equation}
\label{eq55}
B = M \left[b_{0} + k_{4}^{-1}\int{h(t)}dt\right]^{k_{5}},
\end{equation} 
\begin{equation}
\label{eq56}
A = (m M) \left[b_{0} + k_{4}^{-1}\int{h(t)}dt\right]^{k_{5}}.
\end{equation}
Hence, in this case, the metric (\ref{eq1}) reduces to

\[
dS^{2} = - dt^{2} + (m M)^{2}\left[b_{0} + k_{4}^{-1}\int{h(t)}dt\right]^{2k_{5}}dx^{2} + 
\]
\begin{equation}
\label{eq57}
M^{2}\left[b_{0} + k_{4}^{-1}\int{h(t)}dt\right]^{2k_{5}}e^{-2ax} dy^{2} + \left[b_{0} + k_{4}^{-1}
\int{h(t)}dt\right]^{2k_{4}} dz^{2}. 
\end{equation}
The physical parameters $\rho$, $\lambda$, $\rho_{p}$ and the kinematical parameters $\theta$, 
$\sigma^{2}$ for this model are respectively given by 
\begin{equation}
\label{eq58}
8\pi \rho = N(N + 2)\left[\frac{h(t)}{\{b_{0} + k_{4}^{-1}\int{h(t)}dt\}}\right]^{2} 
- \left(\frac{a}{m M}\right)^{2}\left[b_{0} + k_{4}^{-1}\int{h(t)}dt\right]^{-2k_{5}},
\end{equation}
\[
8\pi \lambda = - \left(\frac{a}{m M}\right)^{2}\left[b_{0} + k_{4}^{-1}\int{h(t)}dt\right]
^{-2k_{5}} +
\]
\begin{equation}
\label{eq59}
\frac{N^{2}\Big[8\pi \xi \frac{(2N + 1)}{N^{2}} 
+ \frac{2h_{1}(t)}{N}\{b_{0} + k_{4}^{-1}\int{h(t)}dt\} + h^{2}(t)\left(3 - \frac{1}
{k_{4}k_{5}}\right)\Big]}{[b_{0} + k_{4}^{-1}\int{h(t)}dt]^{2}},
\end{equation}
\begin{equation}
\label{eq60}
8\pi \rho_{p} = \frac{N\Big[2(2 + \frac{1}{N})(N h^{2}(t) - 4\pi \xi) + 2h_{1}(t)
[b_{0} + k_{4}^{-1}\int{h(t)}dt] - \frac{N h^{2}(t)}{k_{4}k_{5}} 
\Big]}{[b_{0} + k_{4}^{-1}\int{h(t)}dt]^{2}},
\end{equation}
\begin{equation}
\label{eq61}
\theta = \frac{(2N + 1)h(t)}{[b_{0} + k_{4}^{-1}\int{h(t)}dt]},
\end{equation}
\begin{equation}
\label{eq62}
\sigma^{1}~ ~_{1} = \sigma^{2}~ ~_{2} = \frac{(N - 1)h(t)}{3[b_{0} + k_{4}^{-1}\int{h(t)}dt]},
\end{equation}
\begin{equation}
\label{eq63}
\sigma^{3}~ ~_{3} = \frac{2(1 - N)h(t)}{3[b_{0} + k_{4}^{-1}\int{h(t)}dt]},
\end{equation}
\begin{equation}
\label{eq64}
\sigma^{4}~ ~_{4} = 0, 
\end{equation}
\begin{equation}
\label{eq65}
\sigma^{2} = \frac{(1 - N)^{2}h^{2}(t)}{3[b_{0} + k_{4}^{-1}\int{h(t)}dt]^{2}}.
\end{equation}
We observe that the equation (\ref{eq54}) has a rich structure and admits different choice   
of function $\xi(t)$. We have to choose $\xi(t)$ in such a manner so that Eq. (\ref{eq54}) be 
integrable. Of course, the choice of $\xi(t)$ is quite arbitrary but since we are looking for 
physically viable models of the universe consistent with observations, one can consider the 
suitable exponential, polynomial and sinusoidal form of the function $\xi(t)$ such that 
Eq. (\ref{eq54}) be integrable. 
\section{Conclusion}
We have presented a new class of Bianchi type III string cosmological models in presence and 
absence of bulk viscosity. In our solution, we have obtained (see, Eq.(\ref{eq21})) a relation 
between metric coefficients from our field equations in a natural way, while in previous 
researches many authors (Bali and Dave, 2002; Wang, 2004) have considered as an {\it ad hoc} 
condition to simplify their equations. If we choose $a = -1$, our model (\ref{eq33}), under some 
particular choice of constants, gives the solution of Bali and Dave (2002). In section 4,
we have obtained a general solution by considering the bulk viscosity as function of time $t$. This 
general solution has a rich structure and admits many number of solutions by suitable choice of 
function $\xi(t)$. Here the choice of $\xi(t)$ is quit arbitrary but since we look for physically 
viable models of the universe, one can choose $\xi(t)$ such that Eq. (\ref{eq54}) be integrable. 

It is observed that the bulk viscosity plays significant role in the evolution of the universe.
In presence of bulk viscosity the model represent an expanding, shearing and non-rotating universe 
without the big bang start. But, in the absence of viscosity, the model starts expanding with a 
big bang at $\tau = 0$.   
\section*{Acknowledgments} 
Authors would like to thank the Inter-University Centre for Astronomy and Astrophysics
(IUCAA), Pune, India for providing facility and support where this work was carried out.
The authors also thank to Prof. Raj Bali for fruitful discussion. The authors are thankful to 
anonymous referee(s) for critical remark and constructive suggestions.  

\end{document}